# Efficiency Models for GaN-based Light-Emitting Diodes: Status and Challenges

**Joachim Piprek**

NUSOD Institute LLC, Newark, DE 19714-7204, USA; piprek@nusod.org

**Abstract:** Light emitting diodes (LEDs) based on Gallium Nitride (GaN) have been revolutionizing various applications in lighting, displays, medical equipment, and other fields. However, their energy efficiency is still below expectations in many cases. An unprecedented diversity of theoretical models has been developed for efficiency analysis and GaN-LED design optimization. This review paper provides an overview of the modeling landscape and pays special attention to the influence of III-nitride material properties. It thereby identifies some key challenges and directions for future improvements.

## 1. Introduction

The 2014 Nobel Prize in Physics was awarded jointly to Isamu Akasaki, Hiroshi Amano and Shuji Nakamura *"for the invention of efficient blue light-emitting diodes which has enabled bright and energy-saving white light sources"* [1]. Their demonstration of GaN-based light-emitting diodes (LEDs) triggered intense worldwide research and development efforts, not only for general lighting applications, but also in many other areas, such as displays, sensing, biotechnology, and medical instrumentation.

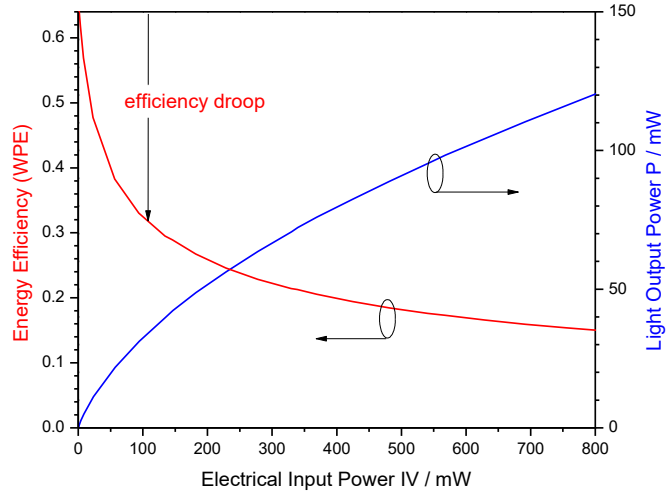

**Figure 1.** Illustration of the energy efficiency (wall-plug efficiency WPE) as ratio of output power to input power.

The promise of superior energy efficiency is the main driving force of many research activities on GaN-LEDs [2,3]. However, high efficiency is only observed at low injection current density and low power (Fig. 1). With rising current, injected electron-hole pairs disappear increasingly in parasitic processes without generating light, thereby causing severe efficiency droop [4]. Still debated is the specific non-radiative mechanism that dominates this efficiency droop, which may be different in different devices. The two leading explanations are Auger recombination inside the light-generating InGaN quantum wells (QWs) [5] and electron leakage from the QWs [6], respectively, in possible combination with other effects. However, very few direct measurements of either mechanism are published, none of which establishes a dominating magnitude. Most publications on efficiency droop mechanisms base their quantitative claims on modeling and simulation [4, 7, 8]. However, the total energy efficiency is usually of greater importance [9] and it is the focus of this paper.

The energy efficiency of the LED semiconductor chip is equivalent to the so-called wall-plug efficiency (WPE) which gives the ratio of light power P emitted from the chip to electrical power IV injected into the chip (I – injected electron-hole current, V - bias) as illustrated in Fig. 1. Different energy loss mechanisms reduce the WPE, which are distinguished by splitting the WPE into separate components:

$$\text{WPE} = \text{ELE} \times \text{EQE} = \text{ELE} \times \text{IQE} \times \text{LEE} = \text{ELE} \times \text{IE} \times \text{RE} \times \text{LEE}. \tag{1}$$

First, the injected electrons lose some energy on their way to the QWs, which is accounted for by the electrical efficiency ELE = $h\nu/qV$ ($h\nu$ - photon energy, q – electron charge). The remaining external quantum efficiency EQE = WPE / ELE is the ratio of emitted photon number to injected number of electron-hole pairs. The conversion of electron-hole pairs into emitted photons is accompanied by carrier losses and by photon losses: EQE = IQE × LEE. The light extraction efficiency LEE accounts for photon losses due to internal light reflection and absorption. The internal quantum efficiency IQE is the fraction of the total current that contributes to the desired photon generation inside the QWs. It can be further separated into the injection efficiency IE (current fraction that enters the QWs) and the radiative efficiency RE (fraction of QW carriers that recombines radiatively).

While WPE can be measured, the analysis of energy loss mechanisms depends mainly on modeling and simulation. For more than a decade, various models have been published for each GaN-LED efficiency component. Emphasizing the influence of material properties, we separate these models in the following into carrier transport models, QW recombination models, light extraction models, and self-heating models.

## 2. Carrier Transport Models

Drift-diffusion models based on the semiconductor transport equations are commonly employed for simulating the carrier movement of electrons and holes in GaN-LEDs [10, 11]. Carrier mobilities are crucial material properties in such models, besides recombination coefficients which are covered in the next section. Together with the free carrier density, the mobility determines the conductivity of each semiconductor layer. The low hole conductivity of Mg-doped III-nitride semiconductors typically dominates the LED bias. Incomplete Mg acceptor ionization is an important but often neglected aspect of drift-diffusion models [12]. Due to the large Mg acceptor ionization energy, high Mg doping densities are required which in turn limit the free hole mobility by impurity scattering. Advanced models for carrier transport parameters have been developed [13]; however, the material quality of fabricated devices is often best represented by experimental data, especially in the case of alloy layers. The semiconductor-metal contact may also have a strong impact on the measured bias which is hard to predict [14].

Several groups employ Monte-Carlo transport models to track the path of individual carriers utilizing tailored scattering models, partially in combination with drift-diffusion models [15,16]. In particular, the movement of high-energy (hot) electrons has been investigated this way.

Quantum mechanical transport models based on the Non-Equilibrium Green's Function (NEGF) method have been published more recently [17, 18]. Such models are especially valuable in the investigation of tunneling and carrier leakage processes. However, the inclusion of electron-hole recombination is difficult and still under development [19]. Simplified tunneling models have been implemented in drift-diffusion simulations to investigate multi-quantum barriers [ 20 ], trap-assisted interband tunneling [21], or LED structures with tunnel-junction cascaded active regions [22, 23].

In an organized effort to demystify the efficiency droop, different transport models were applied to the same experimental LED structure [24], including the common drift-diffusion concept [20], the Monte-Carlo method [25], the NEGF method [17], a ballistic transport model [26], and percolation transport considering random alloy fluctuations [27]. While the normalized efficiency droop was fairly well reproduced in all cases, its physical interpretation is quite different. Some authors conclude dominant Auger recombination, others observe significant electron leakage. Figure 2 shows different current-voltage (IV) curves calculated for this blue LED with a measured turn-on bias of 2.6V [24]. The highest calculated turn-on bias of 3.5V is obtained by the Monte-Carlo model. The common drift-diffusion model gives a turn-on bias of 3.2V. The percolation model results in a soft turn-on starting at 2.8V because carriers search along each interface for the lowest energy barrier. Only the ballistic transport model accounting for high-energy electrons allows for a close fit of the measured IV characteristic. Trap-assisted tunneling was not included in this comparison, which is also known to lower the turn-on bias [21]. However, such IV discrepancies may be caused by the employment of different material parameters. A key parameter is the Mg acceptor density inside the Mg-doped AlGaN electron blocker layer, as only an unknown fraction of Mg atoms forms AlGaN acceptors. This crucial but largely ignored p-AlGaN doping effect creates much uncertainty in GaN-LED simulations [28]. Current crowding [11] and non-uniform carrier injection into the quantum wells [10] is also often neglected or insufficiently analyzed, as most carrier transport simulations are one- or two-dimensional.

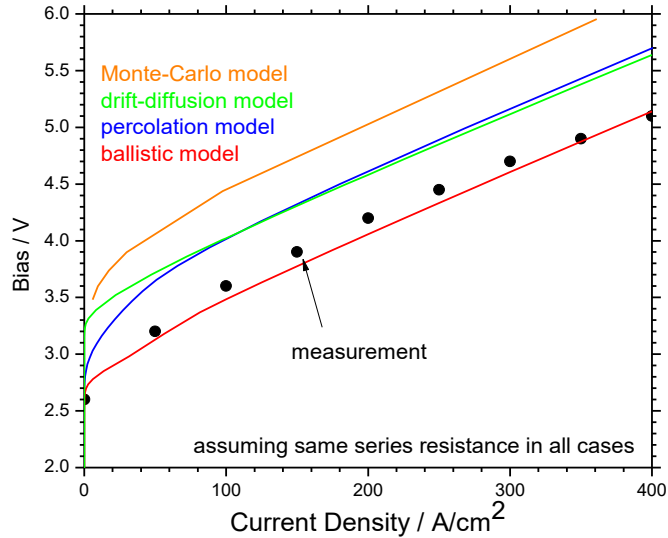

**Figure 2.** Comparison of bias-current characteristics calculated for the same LED structure with different transport models (see text).

The simulated wall-plug efficiency WPE is affected by transport models in two different ways. Firstly, the electrical efficiency ELE depends on the total device bias calculated. Figure 2 demonstrates the bias discrepancy between different modeling approaches. Luckily, the measured device bias can be used in most practical cases to determine the electrical efficiency from the observed photon emission wavelength. In fact, ELE > 1 has been measured on highly optimized GaN-LEDs [29] which is attributed to the absorption of lattice thermal energy by injected carriers before they generate photons [30], encouraging the concept of electroluminescent cooling [31].

Secondly and most importantly, transport models are essential in determining the injection efficiency IE in Eq. (1), i.e., the fraction of carriers that recombines inside the quantum wells, which cannot be measured that easily. Electron leakage into p-doped layers is frequently blamed for the efficiency droop. Such leakage is most often attributed to incomplete carrier capture by the quantum wells [32] or to thermionic emission from the quantum wells [6], and less often to hot electrons [26] or to tunneling [17]. Electrons leaking into the p-doped side of the LED recombine there with holes before those holes can reach the active layers. In other words, electron leakage and reduced hole injection are two sides of the same process. In fact, some authors consider the low hole conductivity of p-doped layers the key reason for the electron leakage [33]. The magnitude of the electron leakage was also found to be highly sensitive to other properties of the AlGaN electron blocker layer (EBL) [34, 35]. Figure 3 plots the relative leakage as function of the built-in polarization and the EBL conduction band offset ratio. Trouble is, both material parameters are not exactly known. Consequently, almost all of the many published simulation studies on EBL design and optimization are quite speculative as long as the leakage current is not validated experimentally. Only very few publication provide such experimental evidence, but none was able to demonstrate that the magnitude of leakage fully explains the magnitude of the efficiency droop [36, 37].

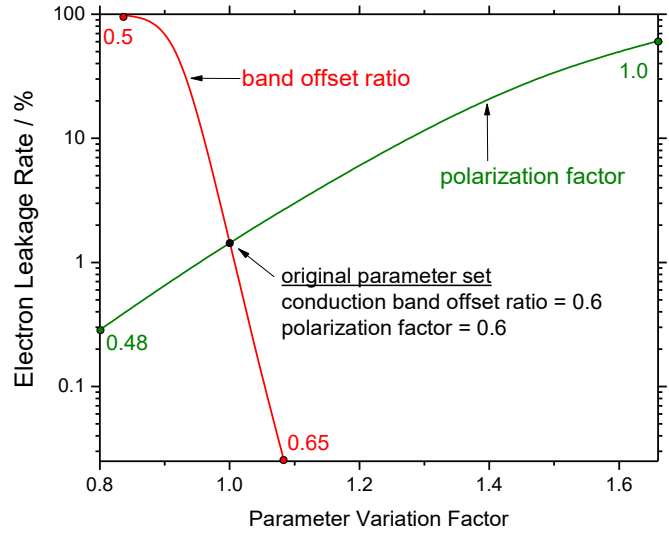

**Figure 3.** The calculated electron leakage is extremely sensitive to variations of band offset and net polarization of the electron blocker layer [35].

## 3. Quantum Well Carrier Recombination Models

Electrons and holes injected into the quantum wells of the LED can be consumed by different recombination mechanisms [38]:

A. crystal defect related recombination
B. radiative recombination
C. Auger recombination

Accordingly, the simple and popular ABC model adds up these different contributions to the total recombination rate $R(n) = A\,n + B\,n^2 + C\,n^3$ (n - QW carrier density; A, B, C – material parameters) and the net current density injected into the QWs $j(n) = q\,d\,R(n)$ (d – total active layer thickness). The radiative efficiency is then given by

$$RE(n) = B\,n^2 / (A\,n + B\,n^2 + C\,n^3). \qquad (2)$$

However, the actual QW carrier density n is typically unknown so that different ABC parameter sets lead to identical efficiency characteristics RE(j) as illustrated in Fig. 4 [39]. In fact, the QW carrier density is known to be non-uniform across a multi-quantum well active region and may even vary inside each QW due to current crowding and/or QW non-uniformities. Various groups proposed modified ABC models, e.g., to account for a reduced active volume [40], inhomogeneous carrier distribution [41], electron leakage [33, 42], photon quenching [43], multi-level defects [44], trap-assisted Auger recombination [45], built-in fields [46], or temperature effects [42, 47, 48]. In any case, ABC models serve as an important bridge between experiment and theory [49]. More detailed models for each of the recombination mechanisms are discussed below.

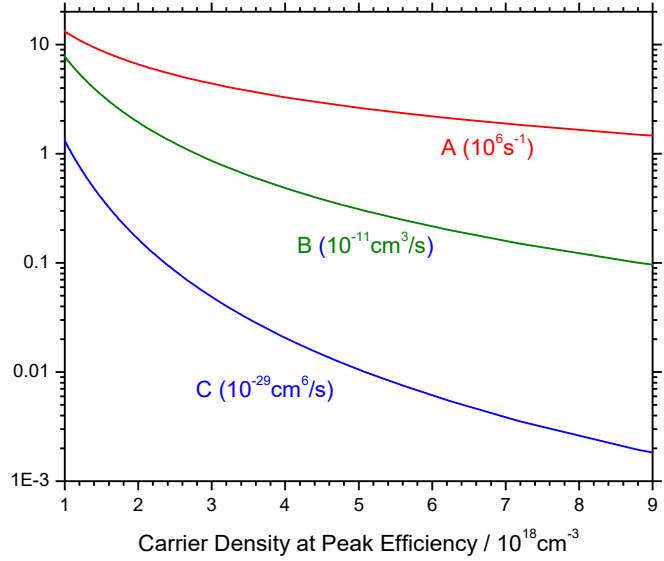

**Figure 4.** Illustration of ABC parameter sensitivity to the quantum well carrier density at peak efficiency. Each vertical combination of recombination coefficients gives identical efficiency vs. current characteristics [39].

*(A) Defect-related Recombination*

The influence of defect-related Shockley-Read-Hall (SRH) recombination on the LED efficiency is undisputed, but it dominates only at low current or in LEDs of poor growth quality with high defect density. Instead of the parameter A in (2), advanced models typically employ SRH lifetimes for electrons and holes as material parameters which can be linked to the density of defects or dislocations [50]. Crystal defects seem unable to cause any efficiency droop since the linear term ($An$) does not increase faster with the carrier density than the light emission ($Bn^2$). For such droop to happen, the A coefficient itself must rise with the defect density in a super-linear way. In other words, the defect-related carrier lifetime needs to decrease rapidly with higher carrier density. Some authors envisioned that QW recombination centers are located on an energy "mountain" so that they can only be reached after the QW "flatland" is filled up with carriers [51]. Other authors put this idea into a numerical model and called it Density Activated Defect Recombination (DADR) [52]. The DADR model shows good agreement with efficiency measurements at low currents, all the way down to very low temperatures. But it fails to reproduce the efficiency droop measured at higher currents. The same is true for a band tail localization model [53] and a droop model based on the influence of QW barrier states [54]. A field-assisted SRH recombination model was proposed to explain the observed temperature sensitivity [55]. However, all these models need to include Auger recombination or electron leakage to fully reproduce droop measurements.

*(B) Spontaneous Recombination (Photon Emission)*

Photon emission from InGaN/GaN quantum wells is handicapped by the built-in polarization field that separates electrons and holes inside the QW (Fig. 5) thereby reducing energy and probability of spontaneous recombination (quantum confined Stark effect) [56]. Advanced

GaN-LED models therefore employ a self-consistent combination of Schrödinger equation and Poisson equation in order to compute the light emission spectrum from the QW energy band structure [10, 57] including various material parameters [58]. The strong electrostatic field is caused by spontaneous and piezo-electric polarization of III-nitride materials grown along the wurtzite c-axis which creates a high density of built-in net charges at all hetero-interfaces (Fig. 6). Various polarization models have been published [59, 60, 61, 62]; however, the predicted polarization charge is typically scaled down in GaN-LED simulations in order to achieve realistic results [14]. A possible reason for this discrepancy is the partial screening of interface polarization charges by other defects. LED growth in different, so-called non-polar or semi-polar crystal directions lowers the polarization field [11, 63, 64, 65]. In any case, the calculated spontaneous emission spectrum often deviates from measurements which may be caused by incorrect predictions of band gap, polarization field, and/or QW structure [10].

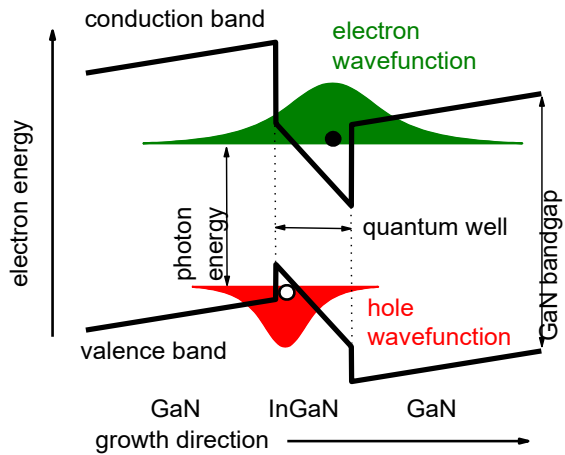

**Figure 5.** Illustration of polarization effects on an InGaN/GaN quantum well in the common Ga-polar growth direction (N-polar growth reverses the polarization field [66]).

Spontaneous recombination saturation effects contribute to the efficiency droop [67] as they change the balance of recombination processes in Eq. (2). Microscopic models reveal that the spontaneous emission rate is proportional to $n^2$ only at low current [68]. At higher current, it may be described by $B=B_0/(1+n/n_0)$ [69]. More recently, photon emission enhancement by the optical LED design is investigated (Purcell effect) [70]. QW coupling with surface plasmons is also predicted to improve photon emission [71, 72]. Nanowire LEDs allow for strain relaxation and enhanced radiative recombination [73]. Bipolar cascade LED designs are envisioned to enable RE > IQE > EQE > 1 at elevated output power as multiple active regions separated by tunnel junctions permit electrons to generate more than one photon [74]. Optical polarization effects gain relevance in AlGaN-based LEDs [75, 76].

*(C) Auger Recombination*

Auger recombination is typically identified as dominating droop mechanism using ABC fits to measured efficiency vs. current characteristics [5]. However, since the $Cn^3$ term in the ABC formula (2) is the only term rising faster with carrier density than the light emission ($Bn^2$), any ABC fit will hold Auger recombination responsible for the droop, no matter what the real cause is. Different models lead to different C-parameter extractions from the same measurement [39]. Figure 7 shows Auger coefficients obtained for various semiconductor materials as a function of the energy band gap. It reveals an uncertainty of several orders of magnitude accompanied by a steep decline with increasing band gap (red symbols). But the nitride data (blue symbols) are clearly outside the broad band predicted which caused early skepticism towards the Auger model for the efficiency droop.

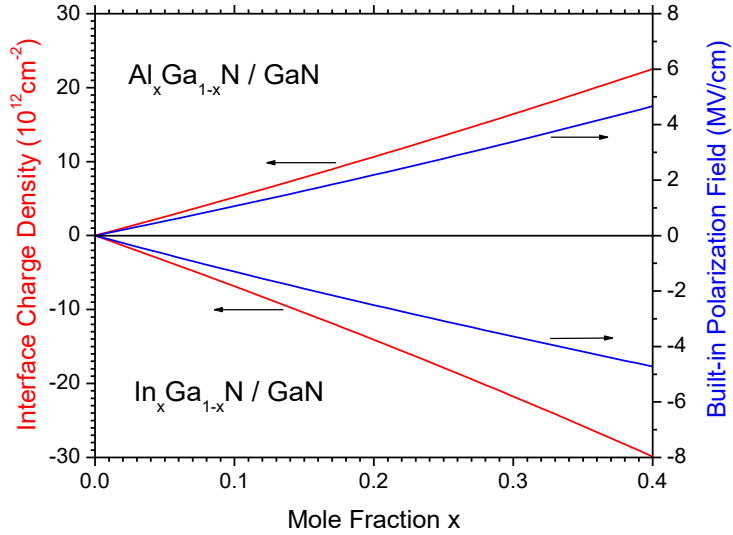

**Figure 6.** Net interface charge density and electrostatic field resulting from built-in III-nitride polarization [77].

Subsequently, several groups have been working on fundamental calculations of the Auger coefficient for III-nitrides. The direct Auger process – involving only three carriers - was initially determined to be very weak [78, 79]. Therefore, indirect Auger recombination was proposed as a possible explanation, which includes electron-phonon coupling and alloy scattering [80, 81]. But the calculated indirect Auger coefficients are only valid for bulk layers and they are below the values required to fully explain the efficiency droop. However, the inclusion of hot Auger electron leakage in the LED model enables lower Auger parameters to cause relevant efficiency droop [82, 83]. Surprisingly, other studies suggest that direct QW Auger recombination may still be strong enough, depending on QW width and composition [84]. More recent Auger recombination models include InGaN alloy disorder [85] or QW carrier localization [86, 87]. Energy band structure theories are the basis of all these models which may include too many approximations and uncertainties to deliver reliable Auger coefficients.

Direct experimental evidence for QW Auger recombination was provided by two somewhat contradicting methods. The first method measured high-energy (hot) electrons emitted from the surface layer of an LED [88]. The authors attribute these hot electrons to the QW Auger process which facilitates electron-hole recombination by transferring the excess energy to a second electron, which thereby becomes "hot" and can travel to the LED surface. Based on Monte-Carlo simulations of this first experiment, other researchers doubt that the Auger-electron can maintain its high energy over such a long travel distance [15]. In fact, the second method assumed a very short travel distance of hot Auger electrons so that they lose their energy quickly and are captured by a neighboring quantum well [89]. But numerical simulations of this second experiment show similar results without Auger recombination [90]. In any case, there is still much uncertainty about the physics of Auger recombination in InGaN QWs and no evidence that this is the only mechanism causing GaN-LED efficiency droop.

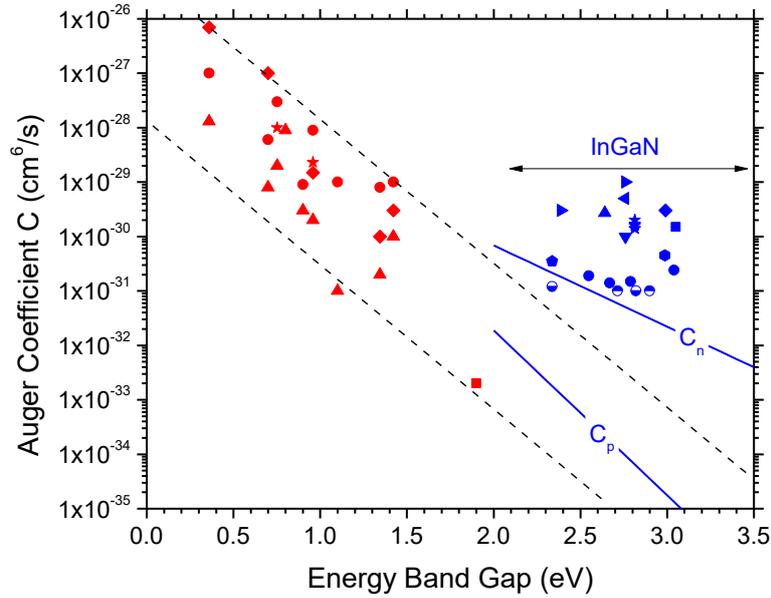

**Figure 7.** Published Auger coefficients for various semiconductors as function of energy band gap. The InGaN data (blue symbols) contradict the steep decline with larger band gap observed with other semiconductors (red symbols). The blue lines are calculated for indirect Auger excitations within conduction bands ($C_n$) or valence bands ($C_p$) of bulk InGaN [81].

**4. Light Extraction Models**

The light extraction efficiency LEE in Eq. (1) often imposes severe limitations on the total energy efficiency WPE. However, this problem captured relatively little attention of the GaN-LED modeling community, because it hardly contributes to the efficiency droop with higher current. Photons generated spontaneously in the active layers travel in all directions inside the LED chip. But only a fraction is able to escape from the chip, due to internal reflection and absorption. Ray tracing models are often employed to calculate LEE [91, 92, 93]. But ray optics fails when structures as small as the photon wavelength are involved. In such cases, Maxwell's equations are usually solved employing the Finite-Difference Time-Domain (FDTD) method, in particular for nano-wire LEDs [94] and photonic-crystal LEDs [95]. Tailored models have been developed for textured LED surfaces [96, 97] and for the influence of Phosphor layers outside the semiconductor chip [93, 98]. Some models also include photon recycling, i.e., their re-absorption by the quantum wells [99]. Light polarization effects need to be considered in deep ultraviolet AlGaN-based LEDs [100].

Refractive index and absorption coefficient are the two key material parameters of LEE models [91], which may also be given as real and imaginary part of the complex dielectric constant [101]. Both depend on material composition and photon wavelength. Based on available measurements, simple refractive index formulas for III-nitride alloys have been developed by several groups [10, 91, 102, 103]. Photon absorption is more difficult to predict as it strongly depends on growth quality and doping [104, 105]. In particular, the high Mg doping density is known to cause significant photon absorption which may be attributed to disorder-induced band tails (Fig. 8) [106].

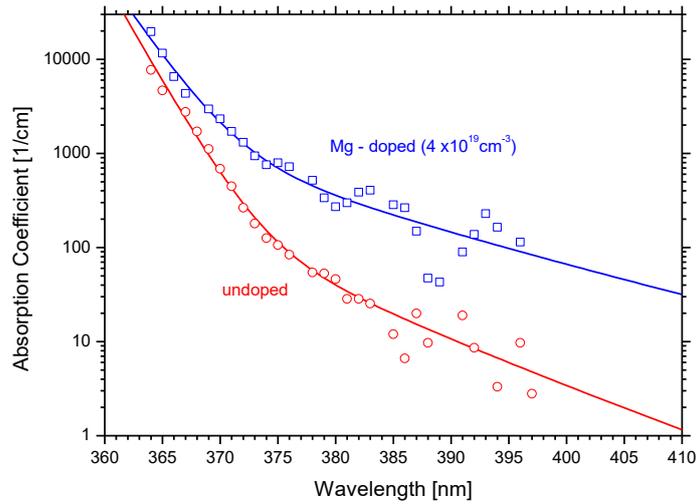

**Figure 8.** GaN absorption coefficients vs. photon wavelength with and without Mg doping. The GaN band gap wavelength is 363nm.   Dots – measurement, lines – fitted band-tail model [106].

## 5. Self-Heating Models

The LED efficiency is known to decline with increasing chip temperature [107, 108,109]. However, self-heating is a three-dimensional problem and only considered by a few published GaN-LED simulations [10, 110, 111]. The thermal conductivity is the key material parameter of such heat flux computations. It is relatively high in perfect GaN crystals, but drops significantly due to phonon scattering at dopants [112], defects [113], and interfaces [114]. Bulk ternary layers suffer from strong alloy scattering of phonons (Fig. 9) [115]. LED chip mounting and packaging also influence the self-heating significantly [116]. Thus, there is much uncertainty about the thermal material properties of real devices so that thermal resistance measurements are often preferred over self-heating simulations.

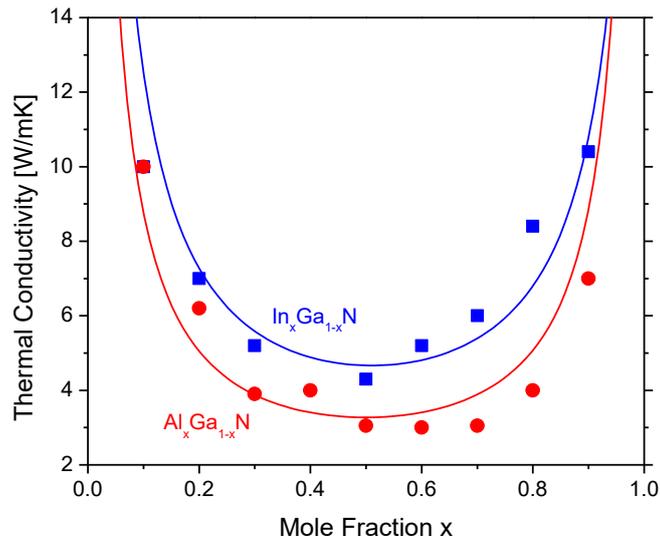

**Figure 9.** Bulk thermal conductivity as function of alloy parameter x as predicted by a molecular dynamics model [117] (dots) and fitted by an analytical function [115]. Measured values for GaN are near 100W/mK.

## 7. Key Modeling and Simulation Challenges

The strong influence of material properties discussed above indicates that the employment of realistic material parameters remains a great challenge for GaN-LED efficiency models. In fact, advanced drift-diffusion simulations of experimental characteristics were shown to validate competing efficiency droop models by simple variation of uncertain parameters [28]. Figure 10 shows good agreement with both efficiency and bias measurements (dots) by enabling dominating carrier loss from Auger recombination (red lines) or from electron leakage (blue lines). The switch was accomplished by changing the Auger coefficient C of the quantum wells and the acceptor doping density $N_A$ of the electron blocking layer, both of which are unknown for real devices. High values produce dominating Auger recombination while low values favor electron leakage in the simulation.

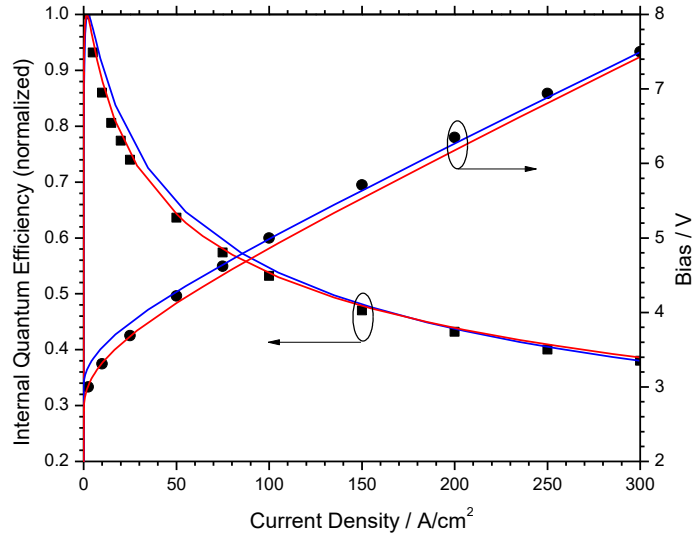

**Figure 10.** Normalized IQE and bias vs. current density (symbols—measurement, red lines—simulation favoring Auger recombination, blue lines—simulation favoring electron leakage) [28].

GaN-LEDs are three-dimensional (3D) objects but most LED simulations are performed in 1D or 2D. Even with uniform material properties in each semiconductor layer, the current flow is often non-uniform in real devices [10, 11], leading to local self-heating, non-uniform carrier density in each QW, non-uniform light emission, and enhanced efficiency droop [118]. While 1D and 2D simulations are very valuable in studying specific mechanisms, they are unable to fully reflect the internal physics and the measured performance of real LEDs.

Another mayor challenge arises from the non-uniform nature of InGaN quantum wells and other thin alloy layers [119]. QWs with low Indium content may exhibit an average Indium atom distance that is larger than the QW thickness. QWs with larger Indium concentration show Indium accumulation regions with lower bandgap, larger free carrier concentration, and stronger Auger recombination. Thus, the typical assumption of uniform QW properties is often invalid. That is why non-uniformity models have been developed in recent years, often embedded in multi-scale LED simulations [119, 120, 121, 122, 123]. However, the more inclusive an LED model is, the more uncertain parameters are usually involved which undermines the reliability of quantitative results.

Artificial intelligence methods also represent a serious challenge. Simulation-based machine learning approaches have been applied to GaN-LED design optimization [124, 125] but produced unreliable results [126]. The great popularity of such methods in materials science [127] and in photonics [128] seems hard to transfer to optoelectronic devices considering their complex internal physics and their material parameter uncertainties. In fact, the strength of machine learning lies in the analysis of large amounts of experimental data which are often routinely collected in the industrial LED production. The combination of reality-trained artificial neural networks (ANNs) with numerical simulations could lead to the creation of realistic digital twins that support the LED design and production process [116, 129, 130].

## 8. Conclusion

Various models have been developed for almost all aspects of GaN-LED device physics which provide valuable insight into internal mechanisms affecting the energy efficiency. However, all models simplify reality so that their relevance and accuracy should always be validated by experiments. Special attention should be paid to the employment of realistic material parameters. Reliable results can only be achieved by interactive and synergetic combination of theoretical modeling, numerical simulation, and experimental investigation.